\documentclass{emulateapj}
\usepackage[utf8]{inputenc}

\begin{document}

\title{Period Variations for the Cepheid VZ Cyg}

\author{
Krittanon Sirorattanakul$^{1}$,
Scott Engle$^{2}$,
Joshua Pepper$^{1}$,
Mark Wells$^{2,3}$,
Clifton D. Laney$^{4,5}$,
Joseph E. Rodriguez$^{6}$,
Keivan G. Stassun$^{7,8}$
}

\affil{$^{1}$Department of Physics, Lehigh University, 16 Memorial Drive East, Bethlehem, PA 18015, USA}
\affil{$^{2}$Department of Astrophysics and Planetary Science, Villanova University, 800 E Lancaster Ave, Villanova, PA 19085, USA}
\affil{$^{3}$Department of Astronomy \& Astrophysics, The Pennsylvania State University, 525 Davey Lab, University Park, PA 16802, USA}
\affil{$^{4}$Department of Physics and Astronomy, Western Kentucky University, 1906 College Heights Blvd, Bowling Green, KY 42101, USA}
\affil{$^{5}$South African Astronomical Observatory, PO Box 9, Observatory, 7935 Cape, South Africa}
\affil{$^{6}$Harvard-Smithsonian Center for Astrophysics, 60 Garden St, Cambridge, MA 02138, USA}
\affil{$^{7}$Department of Physics and Astronomy, Vanderbilt University, 6301 Stevenson Center, Nashville, TN 37235, USA}
\affil{$^{8}$Department of Physics, Fisk University, 1000 17th Avenue North, Nashville, TN 37208, USA}

\begin{abstract}
    The Cepheid Period-Luminosity law is a key rung on the extragalactic distance ladder. However, numerous Cepheids are known to undergo period variations. Monitoring, refining, and understanding these period variations allows us to better determine the parameters of the Cepheids themselves and of the instability strip in which they reside, and to test models of stellar evolution. VZ Cyg, a classical Cepheid pulsating at $\sim$4.864 days, has been observed for over 100 years. Combining data from literature observations, the Kilodegree Extremely Little Telescope (KELT) transit survey, and new targeted observations with the Robotically Controlled Telescope (RCT) at Kitt Peak, we find a period change rate of $dP/dt = -0.0642 \pm 0.0018$ sec yr$^{-1}$. However, when only the recent observations are examined, we find a much higher period change rate of $dP/dt = - 0.0923 \pm 0.0110$ sec yr$^{-1}$. This higher rate could be due to an apparent long-term (P $\approx$ 26.5 yr) cyclic period variation. The possible interpretations of this single Cepheid's complex period variations underscore both the need to regularly monitor pulsating variables, and the important benefits that photometric surveys such as KELT can have on the field. Further monitoring of this interesting example of Cepheid variability is recommended to confirm and better understand the possible cyclic period variations. Further, Cepheid timing analyses are necessary to fully understand their current behaviors and parameters, as well as their evolutionary histories.
\end{abstract}

\section{Introduction}

Classical Cepheids (Cepheids hereafter) are an incredibly useful class of pulsating yellow supergiants. Since the discovery of the Cepheid Period-Luminosity law (the \textit{Leavitt Law} -- \citealt{lea08}) over a century ago, they have become a cornerstone of the Cosmic Distance Scale and a powerful tool for determining the Hubble constant ($H_0$) to a precision of $\sim$1\% \citep{rie16}. 

They also, however, offer valuable insights into stellar astrophysics and the influence that radial pulsations can have on stellar interiors, outer atmospheres, and circumstellar environments. In recent years, multi-wavelength studies of Cepheids have begun detailing new and surprising behaviors. Interferometric studies in the infrared and optical have revealed circumstellar structures around every Cepheid observed to date \citep[see][]{nar16}, in addition to radio observations that have determined mass loss rates of select Cepheids \citep{mat16}. Ultraviolet and far ultraviolet spectra have shown that the outer atmospheres of Cepheids undergo pulsation-phased variations in both emission levels and plasma density \citep{sp82,sp84a,sp84b,boh94,eng09,eng14,eng15,nei16a}. X-ray observations have been used to find stellar companions to Cepheids \citep{eva10,eva16} and to show that the protoype of Classical Cepheids, $\delta$ Cep, is also an X-ray variable \citep{eng17}.  This is all in addition to continuing optical photometry and radial velocity studies attempting to detect the full range of Cepheid variations. Recent efforts have even made use of continuous space-based photometry from satellites such as \textit{CoRoT} \citep{por15}, \textit{BRITE} \citep{smo16}, \textit{MOST} \citep{eva15a} and \textit{Kepler} \citep{der17}.

Once prized for the stability of their pulsations, it has been known for almost a century now that Cepheid pulsation periods can change over time \citep{edd19}. \citet{nei16} give an excellent overview of period variations in not only Cepheids, but other pulsating variable stars as well. Several studies provide a more complete understanding of the types of period variability that Cepheids can display. \citet{sza77,sza80,sza81} are the most comprehensive O-C studies of galactic Cepheids available. Updated O-C data sets for select Cepheids, some of which display potential companion-induced period variations, can also be found in \citet{sza89,sza91}. The important role that amateur observers can play in Cepheid studies is highlighted by \citet{ber03}, who analyzed the AAVSO database and derived numerous times of maximum light for a number of bright Cepheids. And finally, very high cadence and precision radial velocity studies are now being used to also search for period variations and potential companions to bright Cepheids \citep{and16a,and16b}.

Monitoring the pulsations of Cepheids has become a powerful tool for studying stellar evolution on human timescales. Whether a Cepheid's period is increasing or decreasing, and the rate at which it does so, reveals evolutionary changes in the mean density of the star. As a Cepheid evolves towards the cool edge of the instability strip, its overall size grows and thus the density decreases. As the pulsation period of a star is inversely related to its mean density (the Period-Density relationship), the pulsation period increases. Conversely, when a Cepheid evolves toward the hotter edge of the instability strip, its overall size shrinks, density increases and thus the period decreases. The rate of period change can also theoretically be used to determine specifically where a Cepheid is within the instability strip \citep{tur06}.

Normally, monitoring the evolution of a Cepheid's pulsation period requires yearly observing campaigns designed to develop full phase coverage in each year. This can involve a significant investment of telescope time. However, the advent of wide field photometric surveys such as, e.g., the All-Sky Automated Survey \citep[ASAS --][]{Pojmanski:1997} and the Kilodegree Extremely Little Telescope \citep[KELT --][]{Pepper:2007,Pepper:2012}, now provide excellent datasets for carrying out period change studies of a large number of Cepheids spread out across the sky. Here we report on a pilot period study of the Cepheid VZ Cyg using a combination of both survey data and targeted photometric observations.

\section{VZ Cyg}

VZ Cyg (BD+42 4233; $V\approx$ 8.62--9.29; F5--G0 II; $\alpha$ = 21:51:41.44, $\delta$ = +43:08:02.5) was first discovered to be a variable star by \citet{cer04}, using photographic plates obtained by Bla\v{z}ko. Insufficient data were taken at that time to accurately determine the period, though it was noted to be shorter than 5 days. \citet{bla06} later used additional data to populate a better light curve, but one that was plotted with two maxima and two minima. The period was estimated to be 9.727 days. It wasn't until \citet{sea07}, using a much larger dataset of 256 photometric observations, recognized that the earlier reported periods were double the true value which Seares determined as 4.864 days. VZ Cyg is also, as with numerous other Cepheids \citep[see][and references therein]{eva15}, a known spectroscopic binary with a 2183 day orbit ($\sim$5.98 years -- \citealt{gro13}).  

\section{KELT Observations}

The Kilodegree Extremely Little Telescope is a photometric survey using two small-aperture (42 mm) wide-field robotic telescopes, KELT-North at Winer Observatory in Arizona in the United States \citep{Pepper:2007}, and KELT-South at the South African Astronomical Observatory (SAAO) near Sutherland, South Africa \citep{Pepper:2012}. The KELT survey covers over 70$\%$ of the sky and is designed to detect transiting exoplanets around stars in the magnitude range $8 < V < 11$, but can derive photometry for stars between $7 < V < 14$.  It is designed for a high photometric precision of RMS $<1\%$ for bright, non-saturated stars.

VZ Cyg is located in KELT-North field 12, which is centered at J2000 $\alpha =$ 21.4$^{h}$, $\delta =$ +31.7$\degr$. Field 12 was monitored for seven seasons from UT 2007 June 08 to UT 2013 June 14, acquiring a total of 5159 images after post-processing and removal of bad images (see Table \ref{tab:KELT} for KELT photometry of VZ Cyg). Because the KELT-North telescope is located in the American southwest, the monsoon weather prevents observations in the middle of summer (roughly early July to the beginning of September).  Because that is when the visibility of VZ Cyg peaks, our data contain gaps at those times.  The dates and number of images for each observing season are shown in Table \ref{tab:seasons}, with each observing season separated in (a) and (b) segments due to the monsoon gap.

Because KELT uses a German Equatorial Mount, the telescope performs a flip when crossing the meridian, so data acquired in the eastern orientation must be reduced separately from data acquired in the western orientation.  In this analysis, we have combined the east and west KELT light curves for VZ Cyg into a single data set.

\begin{table}
\centering
\caption{KELT Observations of VZ Cyg}
\label{tab:seasons}
\begin{tabular}{ |c|c|c|c| }
\hline
    Season & Start Date & End Date & Number of Images \\
 \hline
 1a & 2007 Jun 08 & 2007 Jun 27 & 435 \\ 
 1b & 2007 Sep 19 & 2007 Oct 13 & 116 \\
 2a & 2008 Apr 24 & 2008 May 21 & 187 \\
 2b & 2008 Sep 18 & 2009 Jan 08 & 813 \\
 3a & 2009 Mar 26 & 2009 Jun 23 & 408 \\
 3b & 2009 Sep 22 & 2009 Dec 20 & 673 \\
 4a & 2010 Apr 26 & 2010 Jun 28 & 375 \\
 4b & 2010 Sep 26 & 2010 Dec 18 & 641 \\
 5a & 2011 Apr 30 & 2011 Jun 17 & 212 \\ 
 5b & 2011 Sep 21 & 2011 Dec 17 & 628 \\
 6a & 2012 Apr 22 & 2012 Jun 22 &  97 \\
 6b & 2012 Sep 17 & 2012 Dec 13 & 464 \\
 7a & 2013 May 07 & 2013 Jun 14 & 110 \\
 \hline
\end{tabular} 
\end{table}

\section{RCT Photometry}

\begin{figure}[hbtp!]
\centering
\includegraphics[width=0.45\textwidth]{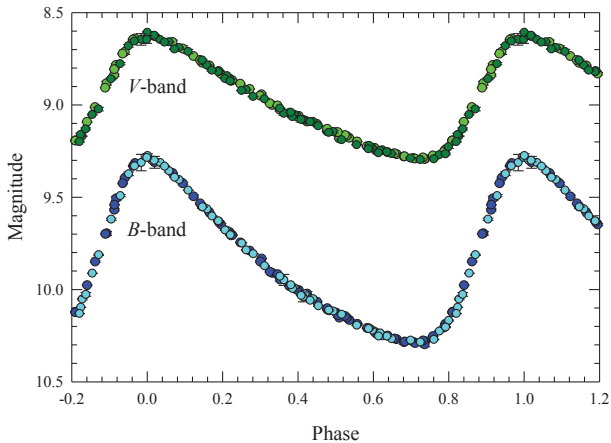}
\caption{\textit{BV} photometry of VZ Cyg obtained with the RCT, phased using a new ephemeris, determined by this study, of: $HJD_{\rm max} = 2457665.6738 + 4.864207~{\rm days} \times E$. Bright green and dark green circles represent the 2015 and 2016 \textit{V}-band data, respectively, and blue and cyan circles represent the 2015 and 2016 \textit{B}-band data, respectively.}
\label{fig:rctphot}
\end{figure}

Recent CCD $BV$ photometry of VZ Cyg (see Table \ref{tab:rct} and Figure \ref{fig:rctphot}) were also carried out for this program with the 1.3m \textit{Robotically Controlled Telescope} (RCT -- \citealt{str14}) at Kitt Peak National Observatory (KPNO). Two seasons were obtained -- Season 1 from 2015 Aug 15 to 2015 Dec 14 and Season 2 began on 2016 Aug 20 to 2016 Nov 19. Observed amplitudes of $A_V$ = 0.680 mag and $A_B$ = 1.004 mag are found, making VZ Cyg a moderate amplitude Cepheid for its pulsation period \citep{kla09}.

\section{Analysis}

\subsection{Outlier Rejection}

To remove spurious data points, we phase the KELT light curve of VZ Cyg based on the reported period from \citet{sza91}, 4.86445 days. We then divide the phased light curve into 25 bins, and for each bin, we compute the median absolute deviation (MAD) of the magnitude and reject any data points that are more than 4 MAD from the median magnitude. We do that once more and reject a total of 76 data points.

\subsection{Periodicity and Blending}
After performing outlier rejection, we run a period search algorithm, analysis of variance \citep[AoV --][]{S-C:1989}, implemented in the VARTOOLS light curve analysis program \citep{Vart:2016}, to determine the Cepheid period from the KELT data. AoV searches for a period by using phase binning. It is sensitive to detecting high-amplitude non-sinusoidal signals. We search a period range of 2 and 20 days with 20 bins, and find a peak at 4.864295 days.

\begin{figure}
\centering
\includegraphics[width=0.45\textwidth]{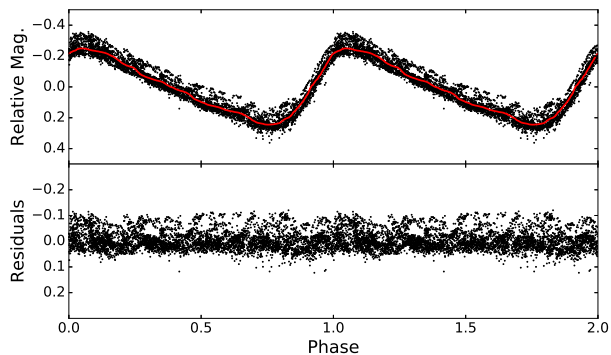}
\caption{{\it Top panel}: KELT light curve of VZ Cyg Cepheid, phased to the initially-determined Cepheid period of 4.864295 days.  Overlaid on top is the median smoothing fit in red.  {\it Bottom panel}: Residuals after subtracting off the model fit.}
\label{fig:ModelFitting}
\end{figure}

\begin{figure}
\centering
\includegraphics[width=0.45\textwidth]{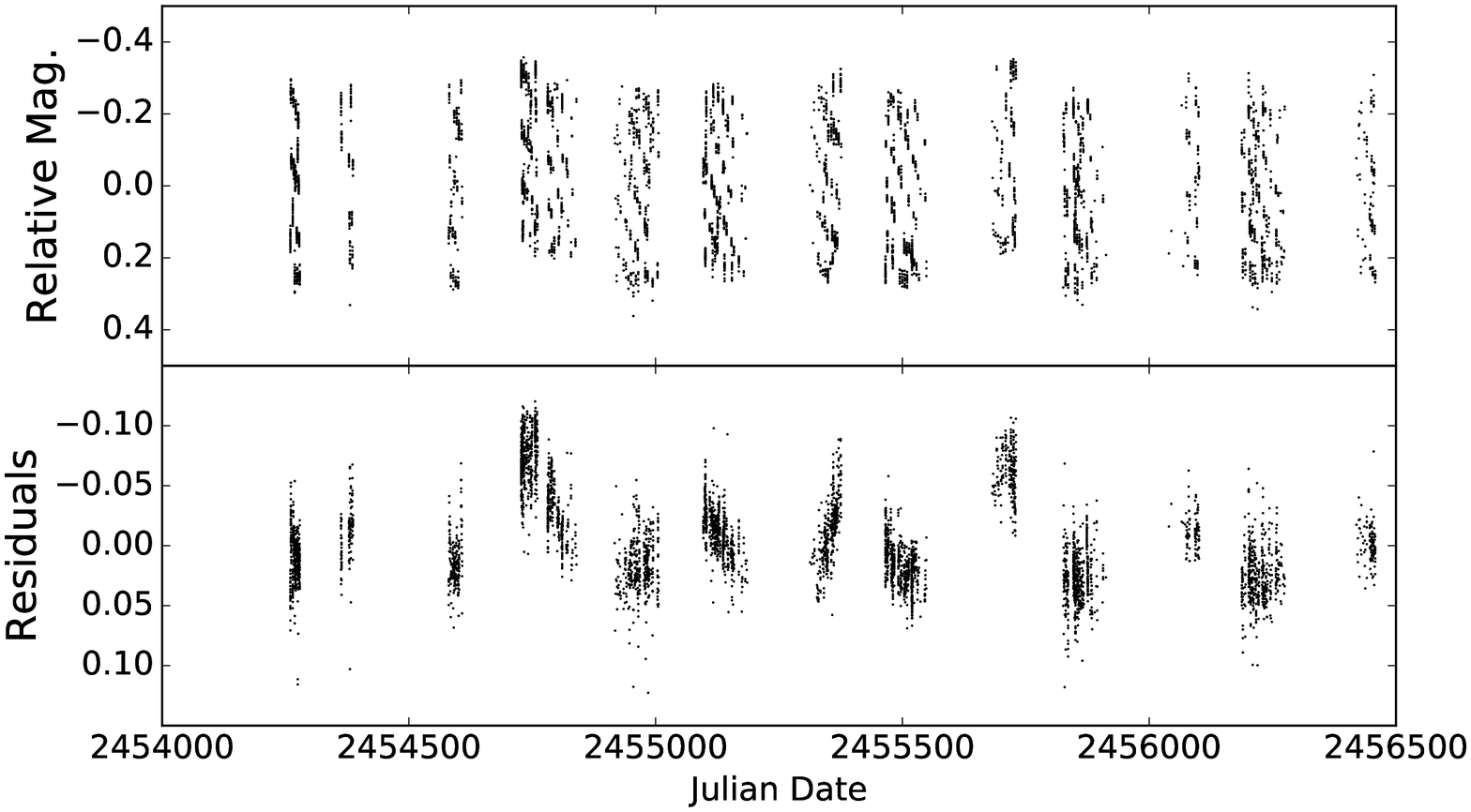}
\caption{{\it Top panel}: Unphased KELT photometry of VZ Cyg. {\it Bottom panel}: Residuals after subtraction of the Cepheid pulsations.}
\label{fig:TimeSeries}
\end{figure}

The phased KELT light curve of VZ Cyg is shown in the top panel of Figure \ref{fig:ModelFitting}, but it appears to show some structure on top of the Cepheid signal.  In order to identify that behavior we model and subtract the Cepheid pulsations and examine the residuals.  We employ a median smoothing fit, using a smoothing length of 1/20th of the pulsational period.  The fit from that smoothing is shown in the top panel of Figure \ref{fig:ModelFitting}, and the residuals between that fit and the data are shown in the bottom panel. Figure \ref{fig:TimeSeries} shows the full unphased KELT light curve, along with the residuals after subtracting the Cepheid pulsations.

\begin{figure}
\centering
\includegraphics[width=0.45\textwidth]{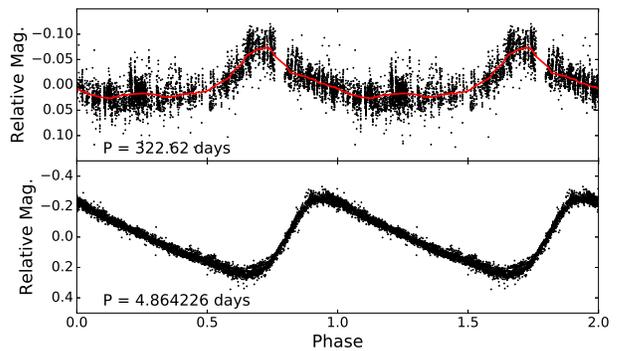}
\caption{{\it Top panel}: Residuals after subtraction of the Cepheid pulsations, phased to 322.62 days, due to blended background Mira variable. Overlaid on top is the median smoothing fit in red. {\it Bottom panel}: Light curve of VZ Cyg after subtraction of the residuals, phased to a new Cepheid period of 4.864226 days.}
\label{fig:ResidualsPhaseDiagram}
\end{figure}

We then search for periodic signals in the residuals after subtracting the Cepheid pulsations. This time we use the generalized Lomb-Scargle \citep[L-S --][]{Press:1992,Zechmeister:2009}, also implemented in VARTOOLS. Generalized L-S searches for a period by fitting sinusoids, and which we have found more sensitive to low amplitudes signals than other algorithms. We search across a period range of 10 and 500 days. L-S finds a peak at 322.62 days. The phased plot of the residuals to that period is shown in the top panel of Figure \ref{fig:ResidualsPhaseDiagram}.  

After re-examining the original KELT images, we have traced the long-term modulation to blending between VZ Cyg and a nearby star V673 Cyg, located at $\alpha$ = 21:51:37.75, $\delta$ = +43:09:58.7, which is 2.14 arcmin from VZ Cyg, or 5.6 KELT pixels.  That star is a known Mira variable, and is somewhat blended into the effective aperture for VZ Cyg in KELT.  We use the median smoothing method, applied to the phased residuals from the Cepheid pulsations shown in the top panel of Figure \ref{fig:ResidualsPhaseDiagram}, to represent the variability of that star, and subtract that signal from the light curve of VZ Cyg.  We then recompute the period of the Cepheid variability, this time using multiharmonic AoV \citep{S-C:1996} using 4 harmonics, finding a period of 4.864226 days, and we display the resulting light curve in the bottom panel of Figure \ref{fig:ResidualsPhaseDiagram}.  We use that final light curve for the analysis described below.

\subsection{Stellar Parameters}

Intensity-weighted mean magnitudes of $\langle B \rangle$ = 9.799 $\pm$ 0.005 and $\langle V \rangle$ = 9.005 $\pm$ 0.005 were derived for VZ Cyg from the RCT photometry. This gives an observed $\langle B \rangle$ -- $\langle V \rangle$ = 0.794, but VZ Cyg has a spectroscopically determined color excess of $E_{B-V}$ = 0.266 \citep{gro13}, resulting in a dereddened color index ($\langle B \rangle$ -- $\langle V \rangle$)$_0$ = 0.528. VZ Cyg was also included in the first \textit{Gaia} data release \citep[\textit{Gaia} DR1]{lin16}, which determined a parallax of $\pi = 0.545 \pm 0.228$ mas. \citet{ast16} showed that applying a Milky Way stellar observability prior resulted in improved DR1 distances for targets nearer than 2 kpc, while an exponentially decreasing stellar density prior worked better for distances larger than 2 kpc. Previous distance estimates for VZ Cyg, e.g. $1849 \pm 139.9$ pc \citep{gro13}, are close enough to the 2 kpc demarcation value that an average of the Milky Way and exponential prior-based distances was deemed to be the best representation of the \textit{Gaia}-determined distance until future data releases become available. Averaging the mode Milky Way and exponential distance values (1692 and 2029 pc, respectively) from \citet{ast16}, and combining their standard errors, gives a distance of $1861 \pm 879$ pc. This value is in good agreement with previous distance estimates, such as that of \citet{gro13}. Using the RCT photometry, the reddening value from the literature, and the \textit{Gaia}-derived distance, we calculate an absolute magnitude of $M_V = -3.17^{+1.39}_{-0.84}$ for VZ Cyg. The errors are large, but this value of $M_V$ is calculated using an early \textit{Gaia} parallax with an appreciable error that will vastly improve over the mission lifetime. However, the absolute magnitude still agrees well with the spectroscopically determined (via Fe \textsc{ii} / Fe \textsc{i} line ratios) value of $M_V = -3.11\pm0.18$ determined by \citet{kov10}.

We note that, at the time of this writing, the {\it Gaia\/} $\pi$ values potentially have systematic uncertainties that are not yet fully characterized but that could reach $\sim$300~$\mu$as\footnote{See \url{http://www.cosmos.esa.int/web/gaia/dr1}.}. Preliminary assessments suggest a global offset of $-0.25$~mas (where the negative sign indicates that the {\it Gaia\/} parallaxes are underestimated) for $\pi \gtrsim 1$~mas \citep{Stassun:2016b}, corroborating the {\it Gaia\/} claim, based on comparison to directly-measured distances to well-studied eclipsing binaries by \citet{Stassun:2016a}. \citet{Gould:2016} similarly claim a systematic uncertainty of 0.12~mas. \citet{Casertano:2017} used a large sample of Cepheids to show that there is likely little to no systematic error in the {\it Gaia\/} parallaxes for $\pi \lesssim 1$~mas, but find evidence for an offset at larger $\pi$ consistent with \citet{Stassun:2016b}. Thus the available evidence suggests that any systematic error in the {\it Gaia\/} parallaxes is likely to be small, and probably negligible for $\pi < 1$~mas. For the purposes of this work, we use and propagate the reported {\it random} uncertainties on $\pi$ only, emphasizing that additional (or different) choices of $\pi$ uncertainties may be applied in the future. 

\section{O-C Analysis}

\begin{figure}
\centering
\includegraphics[width=0.45\textwidth]{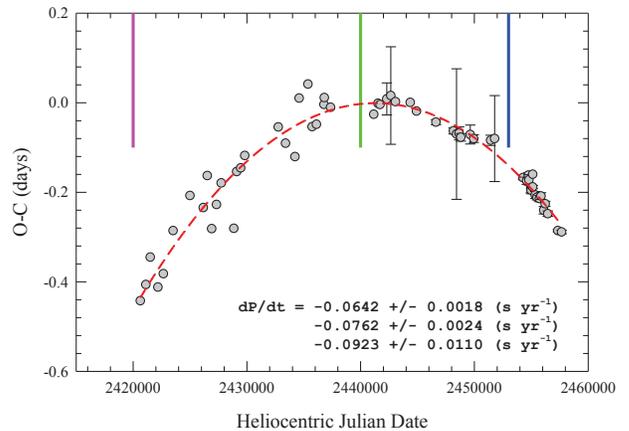}
\caption{The full O-C diagram for VZ Cyg, including all literature data along with those from KELT and the RCT. The three colored vertical lines at the top of the plot demarcate the three time spans of O-C data separately analyzed. At the bottom of the plot, the three rates of period change found from quadratic fits to the data sets are given. From top to bottom, they represent rates for the full data set, the O-C data since HJD = 2440000, and the O-C data since HJD = 2454000, respectively.}
\label{fig:vzcygoc}
\end{figure}

All available times of maximum light for VZ Cyg are compiled from the literature, and combined with the recent maxima from KELT survey photometry and pointed CCD photometry gathered by us with the RCT (see Table \ref{tab:oc}). A Fourier series fit is applied to the combined RCT data \citep[see][]{eng14,eng15}, and the fit results serve as a template light curve for determining times of maximum light and errors for the individual KELT and RCT seasons. However, KELT photometry is not taken through a standard photometric filter, but rather a Kodak Wratten No. 8 red-pass filter \citep{Pepper:2007}. The resulting KELT system response function peaks just short of the standard $R$ bandpass. As a result, adjustments have to be made to the KELT timings, as per the $BVRI$ amplitudes and phase shifts of \citet{free88}. For VZ Cyg, the timing shifts between the $V$ and $R$ photometry of \citet{ber08} were used to calculate the appropriate KELT timing shift of $-0.025 \pm 0.005$ days. Also, in cases where photometric data sets are made available, but no times of maxima are published, these times will also be determined by fitting our template curve to the data. This is known as the Hertzsprung method \citep{her19}. Previously published times of maximum were assigned weights, but no errors. Therefore, when fitting the O-C data, the weights were taken into account so that all data could be handled equally. We compute O-C data for VZ Cyg using the ephemeris of \citet{sza91}:
\begin{center}
$HJD_{\rm max} = 2441705.702 + 4.86445~{\rm days} \times E$
\end{center}
where \textit{C} is the computed time of maximum light, and \textit{E} is the epoch of the observation. The O-C data are presented in Table \ref{tab:oc} and Figure \ref{fig:vzcygoc}. As the figure shows, the pulsation period of VZ Cyg is continually decreasing over time, as also reported by \citet{tur98}. However, the situation becomes more complex depending on which time span of data is analyzed. Analyzing the complete O-C data set returns a rate of period change dP/dt = $-0.0642 \pm 0.0018$ sec yr$^{-1}$, which is considerably slower than the rate of $-0.2032$ sec yr$^{-1}$ reported by Turner. Such a difference can be understood, though, given the dynamic nature of Cepheid evolution and the fact that the current study benefits greatly from the data published in the nearly two decades that have passed since Turner's analysis. This rate of period decrease places VZ Cyg in the second crossing of the instability strip \citep{tur06}. 

The addition of new data to an O-C diagram provides benefits beyond the extended time span. In particular, for known variable stars like VZ Cyg, earlier data are often visual and usually less precise than more modern photometry, especially from photoelectric or CCD instruments. This means that the scatter of an O-C diagram tends to significantly decrease over time. Therefore, more recent epochs of the O-C data set were analyzed separately to see if any further information about the period variability of VZ Cyg could be gleaned. 

\begin{figure*}
\centering
\includegraphics[width=0.9\textwidth]{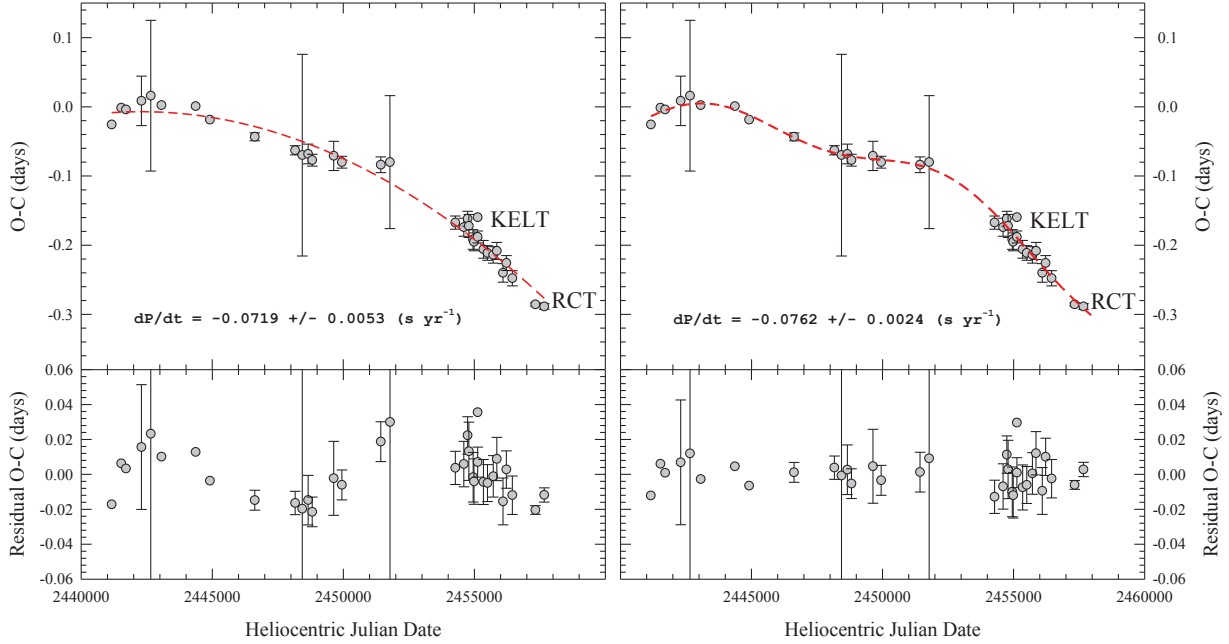}
\caption{The O-C data for VZ Cyg since HJD = 2440000 are plotted. In the left-hand plot, a simple quadratic fit is applied to the data. However, likely due to the increased precision of the photometry and subsequent timings in this time span of data, a potential cyclic O-C variation appears in the residuals of the quadratic fit (bottom left). To improve the fit, a combination quadratic + sinusoidal equation is fitted to the data in the right-hand figure. This fit results in a $\chi^2$ value $\sim2.5\times$ smaller than the quadratic fit, and returns a slightly faster rate of period change ($-0.0762 \pm 0.0024$ s yr$^{-1}$) when compared to either the full O-C data set or the simple quadratic fit, and a cycle length of $26.5 \pm 2.7$ years.}
\label{fig:recent}
\end{figure*}

\begin{figure}
\centering
\includegraphics[width=0.45\textwidth]{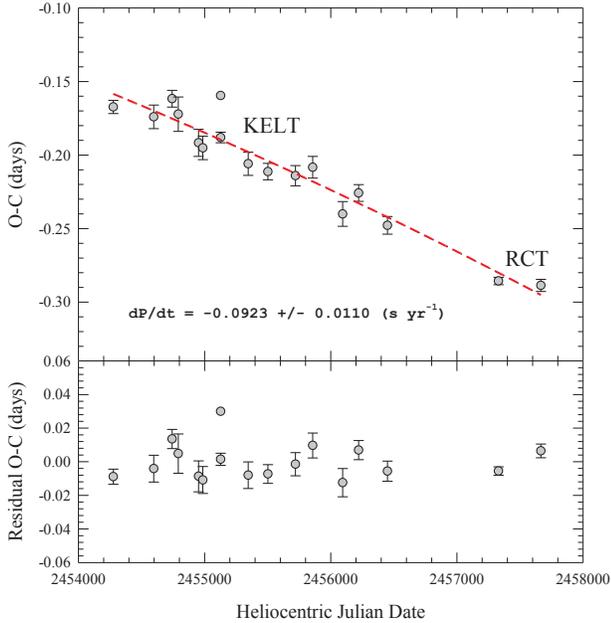}
\caption{The most recent O-C data for VZ Cyg from KELT and the RCT are plotted, along with the quadratic fit which returns a rate of period change of $-0.0923$ s yr$^{-1}$. Residuals to the fit are given in the bottom plot, and the scale of the y-axis is the same as in \ref{fig:recent}. No additional, coherent periodic variations are seen.}
\label{fig:keltrct}
\end{figure}

As shown in (Figure \ref{fig:recent}), when a quadratic fit is applied to only the data after HJD = 2440000, the fit residuals show evidence of a potential cyclic period change. A combination quadratic+sinusoidal fit yields a $\chi^2$ value $\sim2.5\times$ smaller than the quadratic fit, returning a period change rate of dP/dt = $-0.0762 \pm 0.0024$ sec yr$^{-1}$; slightly faster than the period change rate found from the full data set, but still well shy of the value reported by \citet{tur98}. Finally, if only the most recent O-C determinations (those of KELT and the RCT -- Figure \ref{fig:keltrct}) are analyzed, the rate of period change is found to be dP/dt = $-0.0923 \pm 0.0110$ sec yr$^{-1}$. This is the fastest rate of period decrease determined from the O-C data set and subsets, yet still places VZ Cyg in the second crossing of the instability strip. But what to make of the different rates of period decrease? The notion of an accelerating period decrease was considered, but fitting a cubic function to the O-C data set results in a negligible improvement over the quadratic curve, so an accelerating period decrease does not appear to be the case. However, an important issue to account for when dealing with shortened time spans of O-C data is the possibility of additional short-term variations superimposed on any long-term trends. 

Such variations can be seen in the O-C data after HJD = 2440000 (Figure \ref{fig:recent}). The data display an apparent cyclic variability with a period of $26.5 \pm 2.7$ years, determined by a simultaneous quadratic + sinusoidal fit. This additional potential variability further testifies to how complex Cepheid period variations can be. One very exciting explanation for the cycle is the presence of an as yet unknown companion star whose orbit is responsible for the 26.5 year period variations by way of the light travel-time effect (LTTE), sometimes simply called the light time effect (LiTE). Residuals from the most recent O-C data since HJD = 2454000 were also analyzed (Figure \ref{fig:keltrct}), but no firm evidence of coherent cyclic variations is seen. If the data since HJD = 2454000, however, are covering just a small portion of the 26.5 year cyclic period variation, then this would help explain the faster rate of period decrease attributed to these recent timings when compared to either of the larger O-C time spans. 

It is situations like this, when unexpected additional period variations are observed, where the vast potential of automated yearly photometric monitoring of Cepheids becomes clear. As photometry continues to be gathered and analyzed, the full extent of the variations in VZ Cyg but also numerous other variables will become much more apparent.

\section{Evolutionary Results}

With all determined rates of period change placing VZ Cyg in its second crossing of the instability strip, we are given a valuable evolutionary constraint. This allows us to more accurately fit evolutionary tracks and place estimates on certain properties of the Cepheid.

Cepheids and evolutionary tracks have rarely played well together. Cepheid masses determined via evolutionary models were consistently and systematically overestimated when compared to either masses from pulsation models, or for those Cepheids discovered as members of binary star systems. This long-standing problem is referred to as the \textit{Cepheid mass discrepancy}. Amongst the prominent mechanisms put forth to resolved this discrepancy, including convective core overshoot \citep{pra12} and pulsation-enhanced mass-loss \citep{nei08}, is the proper treatment of rotation.

\citet{and14,and17} studied the effects that rotation can have on intermediate mass stellar evolutionary tracks, finding that tracks with rotation effects included can account for the mass discrepancy without needing increased values of core overshoot or mass loss. In short, due to several factors such as rotational mixing bringing additional hydrogen into the core, causing an extended main sequence lifetime and larger resultant helium core, rotation can increase the luminosity of instability strip crossings for a given mass. It also returns larger ages for Cepheids than what have been previously calculated. As discussed in \citet{and14}, the main sequence B-star progenitors of many Cepheids ($M \approx 5 - 9 M_\odot$) typically have rotation rates of $v/v_{crit} \approx 0.3-0.4$, where $v_{crit}$ is the critical rotation velocity.

\begin{figure}
\centering
\includegraphics[width=0.45\textwidth]{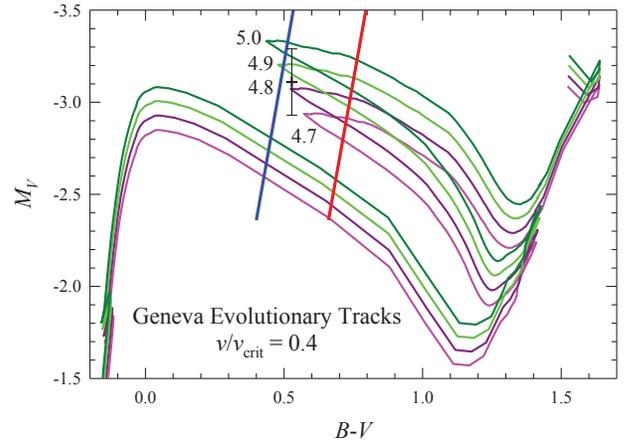}
\caption{Geneva stellar evolutionary tracks \citep{geo13} are plotted, including rotation effects ($v/v_{crit} = 0.4$, where $v_{crit}$ is the critical velocity) along with the location of VZ Cyg and the instability strip boundaries of \citet{tam03}. The tracks are colored as follows: the pink track is $4.7M_\odot$, the purple track is $4.8M_\odot$, the green track is $4.9M_\odot$ and the dark green track is $5.0M_\odot$.}
\label{fig:evol}
\end{figure}

Figure \ref{fig:evol} plots Geneva tracks \citep{geo13} including rotation ($v/v_{crit} = 0.4$) in the region of the instability strip, whose boundaries are taken from Tammann et al. (2003). The location of VZ Cyg is also plotted, according to the values given in Section 4. As the \textit{Gaia}-derived distance still has large errors, the absolute magnitude used for the plot is the spectroscopically determined value from \citet{kov10}. As Figure \ref{fig:evol} shows, VZ Cyg is a good fit for a 4.7--5.0$M_\odot$ star in its second crossing of the instability strip. In looking at the plot, it would first appear that the 4.7$M_\odot$ blue loop doesn't extend to hot enough temperatures to account for VZ Cyg. However, the blue loop lengths are very sensitive to the rotation rates used, so we still consider it as a possible fit.

Using the Geneva tracks plotted in Figure \ref{fig:evol}, we determine a mass of $M = 4.85\pm0.2M_\odot$, a radius of $R = 35\pm2R_\odot$, and an age of $\tau=130\pm6$ Myr for VZ Cyg. The evolutionary radius compares with that of $R = 40\pm19R_\odot$, calculated using the \textit{Gaia} distance and the limb-darkened disk diameter of 0.202 mas from \citet{bou17}. Again, as with the absolute magnitude calculated in Section 4, the radius errors are dominated by those of the parallax. The evolutionary age, as is a known consequence of rotation, is older than other determined values of 71 Myr (\citealt{ach12}; period-age relation of \citealt{bon05}) and 113 Myr (\citealt{mar13}; period-age relation of \citealt{efr03}), though we note that the period-age relation used for the latter age determination is based on LMC cepheids.

\section{Conclusion}

The Classical Cepheid VZ Cyg presents interesting and complex period variations. As it was first classified as a variable star over a century ago \citep{cer04}, VZ Cyg has an excellent timeline of literature observations, which we have combined with data from the literature, the AAVSO archive \citep{aavso}, the KELT survey, and the RCT at Kitt Peak. An O-C analysis of the full data set returns a period change rate of dP/dt = $-0.0642\pm0.0018$ sec yr$^{-1}$. However, recent data indicate faster rates of period decrease. The data after HJD = 2440000 show a period decrease of dP/dt = $-0.0762 \pm 0.0024$ sec yr$^{-1}$, and if only the most recent data (after HJD = 2454000) are analyzed, the rate of period decrease is dP/dt = $-0.0923 \pm 0.0110$ sec yr$^{-1}$.

In addition to the long-term period decrease, quadratic fit residuals from the O-C data since HJD = 2440000 show evidence of a cyclic period variation superimposed on the long-term decrease. This additional cyclic variability has a period of $\sim$26.5 years, but will require further monitoring to confirm. It is likely that the rapid period decrease determined by fitting the most recent O-C data is a result of the influence of the shorter-term period variations. All things considered, the rate of period decrease determined from the O-C data after HJD = 2440000 ($-0.0762$ s yr$^{-1}$) is likely the true rate for VZ Cyg, as this data benefits from the higher precision of more modern photometric methods and instruments, and also covers a long enough time span for the potential cyclic period variations to have been properly accounted for. This value places VZ Cyg in the second crossing of the instability strip. 

Knowing which instability strip crossing VZ Cyg is currently undergoing presents an excellent evolutionary constraint. VZ Cyg was compared to Geneva stellar evolutionary tracks including rotation effects ($v/v_{crit} = 0.4$), wich \citet{and14,and17} have shown can give a more accurate representation of Cepheids, yielding a mass of $M = 4.85\pm0.2M_\odot$, a radius of $R = 35\pm2R_\odot$ and an age of $\tau=130\pm6$ Myr for the Cepheid. Combined with the proper evolutionary tracks, knowing which crossing of the instability strip a Cepheid is in allows for an accurate determination of stellar parameters and a valuable comparison against those derived via other non-evolutionary means. 

The KELT dataset has helped to offer new insights into the variability of VZ Cyg. KELT offers a rich photometric dataset: $\sim$6.5 years of KELT data are analyzed in this paper, with VZ Cyg being observed for $\sim$4--5 months each year, resulting in over 5000 data points. VZ Cyg is just the first target in our program. The final goal is to evaluate the period changes present in all Cepheids that fall within KELT fields, and carry out pointed follow-up photometry of many of these targets with the RCT. Although it is the first target of this study, VZ Cyg immediately shows the large potential of this (and any similar) program. Additional variations in the period of VZ Cyg are now being observed thanks in part to the KELT dataset. Our hope is to significantly improve the understanding of galactic Cepheid period variations on numerous timescales, making use of the yearly KELT observations, along with other available survey data such as the All-Sky Automated Survey (ASAS -- \citealt{Pojmanski:1997}).

With the numerous all-sky (or most-of-the-sky) surveys that have either previously observed, are currently observing, or are planned to begin observing in the near future, it appears likely that our understanding of the behavior and complexity of Cepheids is about to grow considerably.

\acknowledgments

K.S. acknowledges support from Royal Thai Government scholarship. K.G.S. acknowledges partial support from NSF PAARE grant AST-1358862.  We acknowledge with thanks the variable star observations from the AAVSO International Database contributed by observers worldwide and used in this research. This research has made use of the SIMBAD database \citep{Wenger:2000} operated at CDS, Strasbourg, France, the VizieR catalogue access tool, CDS, Strasbourg, France \citep{Ochsenbein:2000}, and NASA's Astrophysics Data System Bibliographic Services. Work performed by J.E.R. was supported by the Harvard Future Faculty Leaders Postdoctoral fellowship.  We thank the anonymous referee for several excellent comments and suggestions that improved the overall clarity and results of the paper.

\facility{AAVSO},\facility{ASAS},\facility{HIPPARCOS},\facility{KELT},\facility{KPNO:RCT},\facility{NSVS}

\bibliographystyle{apj}

\bibliography{main}

\newpage
\begin{table}
\centering
\caption{KELT Photometry of VZ Cyg}
\label{tab:KELT}
\begin{tabular}{ |c|c|c| }
\hline
    HJD & \multicolumn{1}{|p{2cm}|}{\centering KELT relative \\ instr. mag.} & \multicolumn{1}{|p{2cm}|}{\centering Formal \\ mag. error} \\
 \hline
2454259.836392 & 0.160 & 0.004 \\
2454259.840900 & 0.170 & 0.004 \\
2454259.845417 & 0.179 & 0.005 \\
2454259.849932 & 0.162 & 0.005 \\
2454259.854446 & 0.174 & 0.004 \\
2454259.858951 & 0.165 & 0.005 \\
2454259.863474 & 0.167 & 0.004 \\
2454259.867989 & 0.136 & 0.005 \\
2454259.872503 & 0.185 & 0.004 \\
2454259.877008 & 0.158 & 0.005 \\
 \hline
\end{tabular}
\flushleft
\footnotesize{\flushleft This table is available in its entirety in a machine-readable form in the online journal.}
\end{table}

\begin{table}
\centering
\caption{RCT \textit{BV} Photometry of VZ Cyg}
\label{tab:rct}
\begin{tabular}{|l|l|l||l|r|l|}
\hline
HJD          & V-mag  & error  & HJD          & B-mag   & error  \\
\hline
2457249.846954 & 9.1807 & 0.0023 & 2457249.848850 & 10.1530 & 0.0011 \\
2457251.830519 & 8.7829 & 0.0063 & 2457251.832419 & 9.5062  & 0.0119 \\
2457252.830264 & 8.7349 & 0.0019 & 2457252.832166 & 9.4934  & 0.0070 \\
2457253.844356 & 9.0002 & 0.0015 & 2457253.846248 & 9.9104  & 0.0032 \\
2457254.841106 & 9.1779 & 0.0025 & 2457254.842999 & 10.1652 & 0.0034 \\
2457255.821089 & 9.2917 & 0.0020 & 2457255.822977 & 10.2802 & 0.0023 \\
2457263.795291 & 9.0428 & 0.0022 & 2457263.797187 & 9.9823  & 0.0021 \\
2457264.797109 & 9.2239 & 0.0038 & 2457264.799002 & 10.2090 & 0.0021 \\
2457265.807001 & 9.2426 & 0.0006 & 2457265.808890 & 10.1873 & 0.0023 \\
2457269.776281 & 9.2458 & 0.0079 & 2457269.778173 & 10.2370 & 0.0009 \\
\hline
\end{tabular}
\flushleft
\footnotesize{\flushleft This table is available in its entirety in a machine-readable form in the online journal.}
\end{table}

\begin{table}
\centering
\caption{O-C Data for VZ Cyg}
\label{tab:oc}
\begin{tabular}{|l|r|r|c|l|l|}
\hline
$T_{Max}$ (HJD) & Epoch & O-C (days) & Weight & Error & Source       \\
\hline
2420627.5980 & -4333 & -0.4421 & 1 &        & \citet{sza91}   \\
2421114.0790 & -4233 & -0.4062 & 1 &        & \citet{sza91}   \\
2421498.4320 & -4154 & -0.3447 & 1 &        & \citet{sza91}   \\
2422179.3880 & -4014 & -0.4117 & 1 &        & \citet{sza91}   \\
2422656.1340 & -3916 & -0.3818 & 1 &        & \citet{sza91}   \\
2423507.5090 & -3741 & -0.2855 & 1 &        & \citet{sza91}   \\
2424996.1090 & -3435 & -0.2072 & 1 &        & \citet{sza91}   \\
2426163.5500 & -3195 & -0.2343 & 1 &        & \citet{sza91}   \\
2426513.8620 & -3123 & -0.1626 & 1 &        & \citet{sza91}   \\
2426898.0350 & -3044 & -0.2812 & 1 &        & \citet{sza91}   \\
2427321.2960 & -2957 & -0.2273 & 2 &        & \citet{sza91}   \\
2427739.6870 & -2871 & -0.1791 & 2 &        & \citet{sza91}   \\
2428848.6800 & -2643 & -0.2806 & 1 &        & \citet{sza91}   \\
2429096.8940 & -2592 & -0.1536 & 1 &        & \citet{sza91}   \\
2429452.0070 & -2519 & -0.1454 & 1 &        & \citet{sza91}   \\
2429812.0040 & -2445 & -0.1177 & 1 &        & \citet{sza91}   \\
2432755.0600 & -1840 & -0.0540 & 1 &        & \citet{sza91}   \\
2433387.4020 & -1710 & -0.0905 & 1 &        & \citet{sza91}   \\
2434219.1930 & -1539 & -0.1205 & 1 &        & \citet{sza91}   \\
2434589.0220 & -1462 & 0.0103  & 1 &        & \citet{sza91}   \\
2435362.5010 & -1303 & 0.0418  & 1 &        & \citet{sza91}   \\
2435732.1040 & -1228 & -0.0534 & 3 &        & \citet{sza91}   \\
2436106.6720 & -1151 & -0.0481 & 3 &        & \citet{sza91}   \\
2436773.1460 & -1014 & -0.0037 & 3 &        & \citet{sza91}   \\
2436802.3480 & -1007 & 0.0116  & 3 &        & \citet{sza91}   \\
2437352.0090 & -895  & -0.0102 & 3 &        & \citet{sza91}   \\
2441160.8580 & -112 & -0.0256 & 2 &        & \citet{sza91}               \\
2441525.7160 & -37  & -0.0014 & 2 &        & \citet{sza91}               \\
2441705.6980 & 0    & -0.0040 & 3 &        & \citet{sza91}               \\
2442299.1736 & 123  & 0.0087  & 1 & 0.0357 & AAVSO \citep{aavso}         \\
2442654.2859 & 196  & 0.0161  & 1 & 0.1089 & AAVSO \citep{aavso}         \\
2443062.8860 & 280  & 0.0024  & 2 &        & \citet{sza91}               \\
2444366.5570 & 548  & 0.0008  & 3 &        & \citet{sza91}               \\
2444911.3560 & 659  & -0.0185 & 3 &        & \citet{sza91}               \\
2446623.6177 & 1011 & -0.0433 & 3 & 0.0057 & \citet{ber08}               \\
2448160.7643 & 1327 & -0.0628 & 2 & 0.0067 & Hipparcos \citep{hipparcos} \\
2448433.1665 & 1383 & -0.0698 & 1 & 0.1458 & AAVSO \citep{aavso}         \\
2448656.9329 & 1429 & -0.0681 & 2 & 0.0142 & Hipparcos \citep{hipparcos} \\
2448812.5863 & 1461 & -0.0771 & 3 & 0.0085 & \citet{bar97}               \\
2449634.6846 & 1630 & -0.0709 & 2 & 0.0212 & \citet{ber08}               \\
2449946.0003 & 1694 & -0.0800 & 3 & 0.0086 & \citet{ber08}               \\
2451429.6538 & 1999 & -0.0837 & 2 & 0.0114 & NSVS \citep{nsvs}           \\
2451779.8981 & 2071 & -0.0799 & 1 & 0.0961 & AAVSO \citep{aavso}         \\
2454275.2735 & 2584 & -0.1673 & 3 & 0.0095 & this study - KELT             \\
2454596.3205 & 2650 & -0.1740 & 3 & 0.0130 & this study - KELT             \\
2454742.2663 & 2680 & -0.1617 & 3 & 0.0107 & this study - KELT             \\
2454790.9003 & 2690 & -0.1722 & 3 & 0.0167 & this study - KELT             \\
2454951.4077 & 2723 & -0.1917 & 3 & 0.0142 & this study - KELT             \\
2454985.4554 & 2730 & -0.1951 & 3 & 0.0130 & this study - KELT             \\
2455126.5600 & 2759 & -0.1595 & 2 &        & \citet{hue11}               \\
2455126.5315 & 2759 & -0.1881 & 3 & 0.0086 & this study - KELT             \\
2455345.4139 & 2804 & -0.2059 & 3 & 0.0129 & this study - KELT             \\
2455501.0710 & 2836 & -0.2112 & 3 & 0.0106 & this study - KELT             \\
2455719.9684 & 2881 & -0.2140 & 3 & 0.0119 & this study - KELT             \\
2455856.1788 & 2909 & -0.2083 & 3 & 0.0124 & this study - KELT             \\
2456094.5051 & 2958 & -0.2400 & 3 & 0.0134 & this study - KELT             \\
2456220.9950 & 2984 & -0.2258 & 3 & 0.0107 & this study - KELT             \\
2456449.6022 & 3031 & -0.2478 & 3 & 0.0110 & this study - KELT             \\
2457330.0298 & 3212 & -0.2856 & 3 & 0.0025 & this study - RCT              \\
2457665.6738 & 3281 & -0.2886 & 3 & 0.0041 & this study - RCT              \\
\hline
\end{tabular}
\end{table}

\end{document}